\documentclass{PoS}
\usepackage{cite}
\usepackage[utf8]{inputenc}
\usepackage{amsmath}
\usepackage{amsfonts}
\usepackage{amssymb}
\usepackage{amsthm}
\usepackage{enumitem}
\usepackage[T1]{fontenc}
\usepackage{dcolumn}
\usepackage{bm}
\usepackage{bbm}
\usepackage{cancel}
\usepackage[ngerman,english]{babel}
\usepackage[pdftex]{graphicx}
\usepackage{tikz}
\usetikzlibrary{calc,shapes,arrows,patterns,snakes,positioning}
\usepackage{tabularx}
\usepackage{booktabs}
\usepackage{tensor}
\usepackage{xcolor}
\usepackage{mathtools}
\usepackage{float}
\usepackage{exscale}
\usepackage{textpos}
\usepackage{scalerel}
\definecolor{darkblue}{rgb}{0,0,.5}
\usepackage{times}
\DeclareMathSizes{10.95}{9.5}{7}{7} 

\makeatletter 
\newsavebox\myboxA 
\newsavebox\myboxB 
\newlength\mylenA 
 
\newcommand*\xoverline[2][0.75]{% 
    \sbox{\myboxA}{$\m@th#2$}%
    \setbox\myboxB\null% Phantom box 
    \ht\myboxB=\ht\myboxA% 
    \dp\myboxB=\dp\myboxA% 
    \wd\myboxB=#1\wd\myboxA% Scale phantom 
    \sbox\myboxB{$\m@th\overline{\copy\myboxB}$}%  Overlined phantom 
    \setlength\mylenA{\the\wd\myboxA}%   calc width diff 
    \addtolength\mylenA{-\the\wd\myboxB}% 
    \ifdim\wd\myboxB<\wd\myboxA% 
       \rlap{\hskip 0.5\mylenA\usebox\myboxB}{\usebox\myboxA}% 
    \else 
        \hskip -0.5\mylenA\rlap{\usebox\myboxA}{\hskip 0.5\mylenA\usebox\myboxB}% 
    \fi}
\makeatother
\numberwithin{equation}{section}

\let\originalleft\left
\let\originalright\right
\renewcommand{\left}{\mathopen{}\mathclose\bgroup\originalleft}
\renewcommand{\right}{\aftergroup\egroup\originalright}
\newcommand{\e}{\operatorname{e}}
\newcommand{\SU}[1]{\operatorname{SU}\left(#1\right)}
\newcommand{\On}[1]{\operatorname{O}\left(#1\right)}
\newcommand{\Un}[1]{\operatorname{U}\left(#1\right)}
\newcommand{\CP}[1]{\operatorname{CP}\left(#1\right)}

\newcommand{\Sph}[1]{\operatorname{S}^{#1}}
\newcommand{\of}[1]{\left(#1\right)}

\newcommand{\sof}[1]{\bigl(#1\bigr)}
\newcommand{\ssof}[1]{(#1)}
\newcommand{\fof}[1]{\left[#1\right]}

\newcommand{\cof}[1]{\left\{#1\right\}}
\newcommand{\bcof}[1]{\biggl\{#1\biggr\}}

\newcommand{\trace}{\operatorname{tr}}

\newcommand{\avof}[1]{\left\langle #1\right\rangle}

\newcommand{\lagrange}{\mathcal{L}}

\newcommand{\ii}{\mathrm{i}}
\newcommand{\idd}[2]{\mathrm{d}^{#2}\,#1}
\newcommand{\dd}{\mathrm{d}}
\newcommand{\DD}[1]{\mathcal{D}\left[#1\right]}

\newcommand{\partd}[2]{\frac{\partial #1}{\partial #2}}

\newcommand{\spartd}[3]{\frac{\partial^{2} #1}{\partial #2 \partial #3}}

\newcommand{\id}{\mathbbm{1}}
\newcommand{\sfrac}[2]{#1/#2}
\newcommand{\sign}{\operatorname{sgn}}
\DeclarePairedDelimiter\abs{\lvert}{\rvert}%
\DeclarePairedDelimiter\norm{\lVert}{\rVert}%
\makeatletter
\let\oldabs\abs
\def\abs{\@ifstar{\oldabs}{\oldabs*}}
\let\oldnorm\norm
\def\norm{\@ifstar{\oldnorm}{\oldnorm*}}
\makeatother

\newcommand{\ssabs}[1]{\abs*{#1}}

\renewcommand*\[{\begin{equation}}
\renewcommand*\]{\end{equation}}
\renewcommand*\bar[1]{\ThisStyle{\xoverline{\SavedStyle #1}}}
\renewcommand*\hat[1]{\widehat{#1}}
\let\oldstackrel\stackrel
\renewcommand*\stackrel[2]{{\scriptstyle\oldstackrel{#1}{#2}}}

\definecolor{emphcol}{RGB}{0,0,0}
\let\oldemph\emph
\renewcommand*\emph[1]{\oldemph{\textcolor{emphcol}{#1}}}
\newlength{\hatchspread}
\newlength{\hatchthickness}
\newlength{\hatchshift}
\newcommand{\hatchcolor}{}
\tikzset{hatchspread/.code={\setlength{\hatchspread}{#1}},
         hatchthickness/.code={\setlength{\hatchthickness}{#1}},
         hatchshift/.code={\setlength{\hatchshift}{#1}},% must be >= 0
         hatchcolor/.code={\renewcommand{\hatchcolor}{#1}}}
\tikzset{hatchspread=3pt,
         hatchthickness=0.4pt,
         hatchshift=0pt,% must be >= 0
         hatchcolor=black}
\pgfdeclarepatternformonly[\hatchspread,\hatchthickness,\hatchshift,\hatchcolor]% variables
   {custom north west lines}% name
   {\pgfqpoint{\dimexpr-2\hatchthickness}{\dimexpr-2\hatchthickness}}% lower left corner
   {\pgfqpoint{\dimexpr\hatchspread+2\hatchthickness}{\dimexpr\hatchspread+2\hatchthickness}}% upper right corner
   {\pgfqpoint{\dimexpr\hatchspread}{\dimexpr\hatchspread}}% tile size
   {% shape description
    \pgfsetlinewidth{\hatchthickness}
    \pgfpathmoveto{\pgfqpoint{\dimexpr0pt}{\dimexpr\hatchspread+\hatchshift}}
    \pgfpathlineto{\pgfqpoint{\dimexpr\hatchspread+0.15pt}{\dimexpr\hatchshift-0.15pt}}
    \ifdim \hatchshift > 0pt
      \pgfpathmoveto{\pgfqpoint{\dimexpr-0.15pt}{\dimexpr\hatchshift+0.15pt}}
      \pgfpathlineto{\pgfqpoint{\dimexpr\hatchshift+0.15pt}{\dimexpr-0.15pt}}
    \fi
    \pgfsetstrokecolor{\hatchcolor}
    \pgfusepath{stroke}
   }

\pgfdeclarepatternformonly[\hatchspread,\hatchthickness,\hatchshift,\hatchcolor]% variables
   {custom north east lines}% name
   {\pgfqpoint{\dimexpr-2\hatchthickness}{\dimexpr-2\hatchthickness}}% lower left corner
   {\pgfqpoint{\dimexpr\hatchspread+2\hatchthickness}{\dimexpr\hatchspread+2\hatchthickness}}% upper right corner
   {\pgfqpoint{\dimexpr\hatchspread}{\dimexpr\hatchspread}}% tile size
   {% shape description
    \pgfsetlinewidth{\hatchthickness}
    \pgfpathmoveto{\pgfqpoint{0pt}{\dimexpr\hatchshift}}
    \pgfpathlineto{\pgfqpoint{\dimexpr\hatchspread+0.15pt}{\dimexpr\hatchspread+0.15pt+\hatchshift}}
    \ifdim \hatchshift > 0pt
      \pgfpathmoveto{\pgfqpoint{\dimexpr\hatchspread-\hatchshift-0.15pt}{\dimexpr-0.15pt}}
      \pgfpathlineto{\pgfqpoint{\dimexpr\hatchspread+0.15pt}{\dimexpr\hatchshift+0.15pt}}
    \fi
    \pgfsetstrokecolor{\hatchcolor}
    \pgfusepath{stroke}
   }

\pgfdeclarepatternformonly[\hatchspread,\hatchthickness,\hatchshift,\hatchcolor]% variables
   {custom vertical lines}% name
   {\pgfqpoint{\dimexpr-2\hatchthickness}{\dimexpr-2\hatchthickness}}% lower left corner
   {\pgfqpoint{\dimexpr\hatchspread+2\hatchthickness}{\dimexpr\hatchspread+2\hatchthickness}}% upper right corner
   {\pgfqpoint{\dimexpr\hatchspread}{\dimexpr\hatchspread}}% tile size
   {% shape description
    \pgfsetlinewidth{\hatchthickness}
    \pgfpathmoveto{\pgfqpoint{\dimexpr\hatchshift}{0pt}}
    \pgfpathlineto{\pgfqpoint{\dimexpr\hatchshift}{\dimexpr\hatchspread+0.15pt}}
    \pgfsetstrokecolor{\hatchcolor}
    \pgfusepath{stroke}
   }

\pgfdeclarepatternformonly[\hatchspread,\hatchthickness,\hatchshift,\hatchcolor]% variables
   {custom horizontal lines}% name
   {\pgfqpoint{\dimexpr-2\hatchthickness}{\dimexpr-2\hatchthickness}}% lower left corner
   {\pgfqpoint{\dimexpr\hatchspread+2\hatchthickness}{\dimexpr\hatchspread+2\hatchthickness}}% upper right corner
   {\pgfqpoint{\dimexpr\hatchspread}{\dimexpr\hatchspread}}% tile size
   {% shape description
    \pgfsetlinewidth{\hatchthickness}
    \pgfpathmoveto{\pgfqpoint{0pt}{\dimexpr\hatchshift}}
    \pgfpathlineto{\pgfqpoint{\dimexpr\hatchspread+0.15pt}{\dimexpr\hatchshift}}
    \pgfsetstrokecolor{\hatchcolor}
    \pgfusepath{stroke}
   }

\title{
\begin{textblock*}{100pt}(316pt,-164pt)
\textnormal{\small \texttt{CERN-PH-TH/2015-296}}
\end{textblock*}
Lattice simulation of the SU(2) chiral model\\ at zero and non-zero pion density}

\ShortTitle{Lattice simulation of the SU(2) chiral model}

\author{\speaker{Tobias Rindlisbacher}\thanks{E-mail: rindlisbacher@itp.phys.ethz.ch}\\
ETH Z\"urich, Institute for Theoretical Physics,\\ Wolfgang-Pauli-Str. 27, CH - 8093 Z\"urich, Switzerland}

\author{Philippe de Forcrand \thanks{E-mail: forcrand@itp.phys.ethz.ch}\\
ETH Z\"urich, Institute for Theoretical Physics,\\ Wolfgang-Pauli-Str. 27, CH - 8093 Z\"urich, Switzerland\\
and\\
CERN, Physics Department, TH Unit,\\ CH-1211 Gen\`eve 23, Switzerland}

\abstract{We propose a flux representation based lattice formulation of the partition function corresponding to the SU(2) principal chiral Lagrangian, including a chemical potential and  scalar/pseudo-scalar source terms. Lattice simulations are then used to obtain non-perturbative properties of the theory, in particular its mass spectrum at zero and non-zero pion density. We also sketch a method to efficiently measure \emph{general} one- and two-point functions during the worm updates.}

\FullConference{The 33rd International Symposium on Lattice Field Theory\\
		14 -18 July 2015\\
		Kobe International Conference Center, Kobe, Japan*}

\begin{document}
\vspace{-5pt}
\section{Introduction}\vspace{-5pt}
In an earlier study of two-flavor lattice QCD coupled to an isospin chemical potential $\mu$ \cite{Rindlisbacher0}, we were interested in the mass spectrum of the theory as function of $\mu$. From saddle-point/effective-potential calculations based on the $\SU{2}$ chiral effective model (leading term in $\SU{2}$ chiral perturbation theory) \cite{Son}, it is known that as soon as the value of $\mu$ exceeds half the pion mass, the chiral condensate should start to rotate into a pion-condensate and thereby spontaneously break the isospin-$\Un{1}$ symmetry, which is accompanied by the appearance of a massless Goldstone mode. By trying to reproduce the predicted mass spectrum from \cite{Son} within our simulations, we noticed that this is not straightforward in the broken symmetry phase. The difficulties come from the fact that in lattice QCD the fermions are already integrated out exactly, and therefore also the flavor symmetry that should spontaneously break. Also, it seemed that the explicit breaking of the symmetry by the introduction of source terms is not sufficient to get the expected mass spectrum.\\
As lattice QCD is rather expensive to simulate and the construction of meson correlators in the presence of symmetry breaking source terms rather complicated, we decided to first study the issue in a toy model. A first candidate was complex $\phi^4$, coupled to a chemical potential, where in the flux-representation formulation \cite{Gattringer} the field is also integrated out exactly. However, to get closer to the situation of isospin QCD, we decided to generalize the flux-representation formulation from \cite{Gattringer} to the $\SU{2}$-chiral-effective model, as will be described in the remainder of this section\footnote{Note that during the LATTICE 2015 conference, where we first presented our derivation of the flux-representation formulation of the $\SU{2}$ principal chiral model, a similar derivation was published in \cite{Bruckmann} for the case of general $\On{N}$ and $\CP{N-1}$ spin-models. Our partition function for the $\SU{2}$ principal chiral model is essentially the same as their $\On{4}$ version, with added source terms.}. We then continue in Sec.~\ref{sec:simtech} with a brief discussion of an improved method to measure various correlators during the worm update and the presentation of results in Sec.~\ref{sec:results}, before summarizing in Sec.~\ref{sec:conclude}.

\subsection{Continuum Action and Discretization}\label{sec:cont}
We start with the simple case of a chiral effective model for low energy two flavour QCD with degenerate quark masses, $m_{u}\,=\,m_{d}$, and an isospin chemical potential $\mu$. The corresponding Minkowski space continuum Lagrangian reads:
\[
\lagrange_{eff}\,=\,-\frac{f^{2}_{\pi}}{4}\trace\fof{\ssof{\partial_{\nu}\Sigma\,-\,\ii\,\mu\,\delta_{\nu,0}\ssof{\sigma_{3}\Sigma\,-\,\Sigma\sigma_{3}}}\ssof{\partial^{\nu}\Sigma^{\dagger}\,-\,\ii\,\mu\,\delta^{\nu,0}\ssof{\sigma_{3}\Sigma^{\dagger}\,-\,\Sigma^{\dagger}\sigma_{3}}}}\,-\,\frac{f^{2}_{\pi}}{4}\trace\fof{\Sigma^{\dagger}\,S\,+\,S^{\dagger}\,\Sigma}\,\label{eq:continuumlagr}
\]
where $f_{\pi}$ is the \emph{pion decay constant}, $\Sigma\,=\,\frac{\pi^{4}}{f_{\pi}}\,\id\,+\,\frac{\ii\,\bar{\pi}\cdot\bar{\sigma}}{f_{\pi}}$ , with $\pi^4=\sign\ssof{f^{2}_{\pi}\,-\,\abs{\bar{\pi}}^{2}}\,\sqrt{\ssabs{f^{2}_{\pi}\,-\,\abs{\bar{\pi}}^{2}}}$ , is an $\SU{2}$ matrix parametrized by the three pion fields $\bar{\pi}=\cof{\pi^1,\pi^2,\pi^3}$, and $\bar{\sigma}=\cof{\sigma_1,\sigma_2,\sigma_3}$ is a vector containing the three Pauli matrices $\sigma_1,\sigma_2$ and $\sigma_3$. $S\,=\,s_{4}\,\id\,+\,\ii\,\bar{s}\cdot\bar{\sigma}$ also is an $\SU{2}$ matrix, where $s_{4}\propto \,m_{\pi}^{2}\propto f_{\pi}\of{m_{u}+m_{d}}$ can be seen as a pion mass term, arising from a finite quark mass, and the $s^i$ serve as sources for the pion fields.\\
The lattice discretization is straightforward: after a Wick-rotation and the introduction of a finite lattice spacing $a$, the action becomes
\[
S=-\kappa\sum\limits_{x}\bcof{\frac{1}{4}\sum\limits_{\nu=1}^{4}\trace\fof{\Sigma_{x}^{\dagger}\e^{\mu\,\sigma_{3}\,\delta_{\nu,4}}\Sigma_{x+\hat{\nu}}\e^{-\mu\,\sigma_{3}\,\delta_{\nu,4}}\,+\,\Sigma_{x}^{\dagger}\e^{-\mu\,\sigma_{3}\,\delta_{\nu,4}}\Sigma_{x-\hat{\nu}}\e^{\mu\,\sigma_{3}\,\delta_{\nu,4}}}+\frac{1}{4}\trace\fof{\Sigma_{x}^{\dagger}\,S_{x}\,+\,S_{x}^{\dagger}\,\Sigma_{x}}}\ ,\label{eq:latticeaction}
\]
where we have introduced the dimensionless coupling $\kappa\,=\,f^{2}_{\pi} a^{2}$. $\Sigma_{x}$ is already dimensionless and to render also $\pi_{x}^{i}$, $S_{x}$ and $\mu$ dimensionless, we redefine $\pi_{x}^{i}/f_{\pi}\rightarrow \pi_{x}^{i}$, $S_{x}\rightarrow a^2\,S_{x}$ and $a\,\mu\rightarrow \mu$.

\subsection{Partition function and flux representation}\label{ssec:partf}
The partition function of the theory is defined as usual by
\[
Z\,=\,\int\DD{\Sigma}\,\e^{-S\fof{\Sigma}}\,,\quad\text{where}\quad \DD{\Sigma}\,=\,\prod\limits_{x}\frac{\dd\pi_{x}^{1}\wedge\dd\pi_{x}^{2}\wedge\dd\pi_{x}^{3}}{\sqrt{1-\abs{\bar{\pi}_{x}}^2}}\ .\label{eq:partf1}
\]
For non-zero chemical potential $\mu$, \eqref{eq:latticeaction} is in general complex, which leads to a \emph{sign problem} when trying to sample \eqref{eq:partf1} by Monte Carlo.\\
To overcome this problem we can follow the strategy used in \cite{Gattringer} to simulate complex $\phi^4$ theory on the lattice in the presence of a finite chemical potential, and derive the \emph{flux representation} of \eqref{eq:partf1}. To do so, we write \eqref{eq:partf1} out in terms of angular variables, $\dd\Sigma_{x}\,=\,\sin^{2}\of{\alpha_{x}}\sin\of{\theta_{x}}\dd\alpha_{x}\wedge\dd\theta_{x}\wedge\dd\phi_{x}$, 
\[
\pi_{x}^{1}\,=\,\sin\of{\alpha_{x}}\sin\of{\theta_{x}}\cos\of{\phi_{x}}\,,\,\pi_{x}^{2}\,=\,\sin\of{\alpha_{x}}\sin\of{\theta_{x}}\sin\of{\phi_{x}}\,,\,\pi_{x}^{3}\,=\,\sin\of{\alpha_{x}}\cos\of{\theta_{x}}\,,\,\pi_{x}^{4}\,=\,\cos\of{\alpha_{x}}\ .
\]
Then we split the Boltzmann factor into separate exponentials and expand each of them in a power series to arrive at the following expression for the partition function:
\begin{multline}
Z\,=\,\sum\limits_{\cof{k,l,\xi,\chi,p,q,n^3,n^4}}\bcof{\prod\limits_{x,\nu}\frac{\kappa^{\ssabs{k_{x,\nu}}+2\,l_{x,\nu}+\xi_{x,\nu}+\chi_{x,\nu}}}{\of{\ssabs{k_{x,\nu}}+l_{x,\nu}}!\,l_{x,\nu}!\,\xi_{x,\nu}!\,\chi_{x,\nu}!}}\bcof{\prod\limits_{x}\frac{\of{\kappa\,s}^{\ssabs{p_{x}}+2\,q_{x}}\,\of{\kappa\,s_{3}}^{n^3_{x}}\,\of{\kappa\,s_{4}}^{n^4_{x}}\,\e^{\ii\,\phi_{s}\,p_{x}}\,\e^{2\mu\,k_{x,4}}}{2^{\sfrac{\of{\ssabs{p_{x}}+2\,q_{x}}}{2}}\of{\ssabs{p_{x}}+q_{x}}!\,q_{x}!\,n^3_{x}!\,n^4_{x}!}\\
\int\limits_{0}^{\pi}\dd\alpha_{x}\int\limits_{0}^{\pi}\dd\theta_{x}\,\sin^{2}\of{\alpha_{x}}\sin\of{\theta_{x}}\,\sof{\frac{1}{\sqrt{2}}\sin\of{\alpha_{x}}\sin\of{\theta_{x}}}^{\ssabs{p_{x}}+2\,q_{x}+\sum\limits_{\nu}\of{\ssabs{k_{x,\nu}}+\ssabs{k_{x-\hat{\nu},\nu}}+2\,\of{l_{x,\nu}+l_{x-\hat{\nu},\nu}}}}\\
\sof{\sin\of{\alpha_{x}}\cos\of{\theta_{x}}}^{n_{x}^{3}+\sum\limits_{\nu}\of{\chi_{x,\nu}+\chi_{x-\hat{\nu},\nu}}}\,\sof{\cos\of{\alpha_{x}}}^{n_{x}^{4}+\sum\limits_{\nu}\of{\xi_{x,\nu}+\xi_{x-\hat{\nu},\nu}}}\int\limits_{0}^{2\pi}\dd\phi_{x}\e^{\ii\phi_{x}\of{p_{x}+\sum_{\nu}\of{k_{x,\nu}-k_{x-\hat{\nu},\nu}}}}},\label{eq:partf3}
\end{multline}
where $s\,\e^{\pm\ii\phi_{s}}=s_{1}\pm \ii\,s_{2}$ and where we have introduced \emph{flux variables} $k_{x,\nu}\in \mathbb{Z}$, $l_{x,\nu},\,\chi_{x,\nu},\,\xi_{x,\nu}\in\mathbb{N}_{0}$ and \emph{monomer variables} $p_{x}\in\mathbb{Z}$, $q_{x},\,n^3_{x},\,n^4_{x}\in \mathbb{N}_{0}$. $k_{x,\nu}$ counts the net charge flowing from site $x$ to site $x+\hat{\nu}$, $l_{x,\nu}$ counts the number of neutral $\pi^{+}$-$\pi^{-}$ pairs and $\chi_{x,\nu}$, $\xi_{x,\nu}$ the number of $\pi^3$ and $\pi^4$ particles respectively moving between the two sites. The monomer variables count the net monomer content of site $x$, and can be non-zero only if the corresponding sources, $s$, $s^3$ or $s^4$, have non-zero values.\\
Carrying out the angular integrals in \eqref{eq:partf3}, using that for non-negative integers $M$, $N$ the identity $\int\idd{\theta}{}\sin^{M}\of{\theta}\,\cos^{N}\of{\theta}\,=\,\textstyle\frac{1+\of{-1}^{N}}{2}\frac{\Gamma\sof{\frac{1+M}{2}}\,\Gamma\sof{\frac{1+N}{2}}}{\Gamma\sof{\frac{2+M+N}{2}}}$ holds, then yields:
\vspace{-5pt}\[
Z\,=\,\sum\limits_{\cof{k,l,\xi,\chi,p}}\,\bcof{\prod\limits_{x}\,K_{x}\of{\kappa}\,\e^{\ii\,\phi_{s}\,p_{x}}\,\e^{2\mu\,k_{x,4}}\,\delta\sof{p_{x}+\sum\limits_{\nu}\ssof{k_{x,\nu}-k_{x-\hat{\nu},\nu}}}\,w\ssof{A_{x},B_{x},C_{x},p_{x};\kappa,s,s_{3},s_{4}}}\ ,\label{eq:partf4}
\]
where $A_{x}=\sum\limits_{\nu}\ssof{\ssabs{k_{x,\nu}}+\ssabs{k_{x-\hat{\nu},\nu}}+2\ssof{l_{x,\nu}+l_{x-\hat{\nu},\nu}}}$, $B_{x}=\sum\limits_{\nu}\ssof{\xi_{x,\nu}+\xi_{x-\hat{\nu},\nu}}$, $C_{x}=\sum\limits_{\nu}\ssof{\chi_{x,\nu}+\chi_{x-\hat{\nu},\nu}}$,\vspace{5pt} $K_{x}\of{\kappa}=\prod\limits_{\nu}\frac{\kappa^{\ssabs{k_{x,\nu}}+2\,l_{x,\nu}+\xi_{x,\nu}+\chi_{x,\nu}}}{\of{\abs{k_{x,\nu}}+l_{x,\nu}}!\,l_{x,\nu}!\,\xi_{x,\nu}!\,\chi_{x,\nu}!}$ and
\[
w\ssof{A,B,C,p;\kappa,s,s_{3},s_{4}}\,=\,\sum\limits_{q,n^{3},n^{4}=0}^{\infty}\frac{\ssof{\kappa\,s}^{\abs{p}+2\,q}\,\ssof{\kappa\,s_{3}}^{n^3}\,\ssof{\kappa\,s_{4}}^{n^4}}{2^{\sfrac{\ssof{\ssabs{p}+2\,q}}{2}}\ssof{\ssabs{p}+q}!\,q!\,n^3!\,n^4!}W\ssof{A+\abs{p}+2\,q,B+n^3,C+n^4}\ ,\label{eq:weightfunc}
\]
with
\vspace{-20pt}\[
W\of{A,B,C}\,=\,\frac{\frac{1+\of{-1}^{C}}{2}\,\frac{1+\of{-1}^{B}}{2}\,\Gamma\sof{\frac{1+C}{2}}\,\Gamma\sof{\frac{1+B}{2}}\,\Gamma\sof{\frac{2+A}{2}}}{2^{\sfrac{\of{2+A}}{2}}\,\Gamma\sof{\frac{4+A+B+C}{2}}}\ .\label{eq:configweight}
\]
Note that since $\sum_{x,\nu}\ssof{k_{x,\nu}-k_{x-\hat{\nu},\nu}}=0$ identically, the delta function constraints in \eqref{eq:partf4} imply $\sum_{x} p_{x}=0$, and therefore $\prod_{x} \e^{\ii \phi_{s} p_{x}}\,=\,1$. Thus, all terms in  \eqref{eq:partf4} are manifestly real and non-negative so that the partition function can now be computed by importance sampling of the flux variables. However, again due to the delta function constraints in \eqref{eq:partf4}, this has to be done by a worm-algorithm for the $k$ and $p$-variables. Also the $\chi$ and $\xi$-variables are best sampled by a worm, due to the \emph{evenness constraints} for $B$ and $C$ shown at the very left of the numerator in \eqref{eq:configweight}. The values of \eqref{eq:weightfunc} can be pre-computed for the domain of arguments required during a simulation at fixed coupling $\kappa$ and with fixed source terms $\of{s, s_{3},s_{4}}$. This results in a significant speed up and a reduction of statistical noise in observables compared to the Monte Carlo sampling of the $q$, $n^3$ and $n^4$-variables.
\vspace{-5pt}
\section{Worm algorithm and general correlators}\label{sec:simtech}\vspace{-5pt}
Our worm algorithm is based on the idea to allow the Markov chain to change between the sampling of the partition function $Z$ in \eqref{eq:partf4} itself and the sampling of the various one- and two-point functions that can be obtained by taking derivatives of $Z$ with respect to the different source terms $s$, $s_3$, $s_4$ (considering them temporarily as local fields). Recording the fractional Monte Carlo time the algorithm spends for example sampling configurations that allow for the presence of an external $\pi^3$ somewhere in the system, then corresponds to a measurement of the expectation value $\avof{\pi^{3}}=\frac{1}{Z}\partd{Z}{s_{3}}$ . Similarly, the fractional MC time the algorithm spends sampling configurations that allow for two external fields, say a $\pi^{+}$ at some site $x$ and a $\pi^{-}$ at another site $y$, corresponds to the expectation value of the point to point correlator $\avof{\pi^{+}\of{x}\pi^{-}\of{y}}=\frac{1}{Z}\spartd{Z}{s^{-}_{x}}{s^{+}_{y}}$ .\\
The transition probabilities in the Markov chain for changing from the sampling of $Z$ to the sampling of one of the one- or two-point functions can be obtained by noting that the derivative of the partition function with respect to, say $s^{\pm}_{z}$, can be written as
\vspace{-5pt}\[
\frac{1}{\kappa}\partd{Z}{s_{z}^{\pm}}=\hspace{-6pt}\sum\limits_{\cof{k,l,\xi,\chi,p}}\bcof{\prod\limits_{x}\,K_{x}\of{\kappa}\e^{\ii\,\phi_{s}\,p_{x}}\e^{2\mu\,k_{x,4}}\,
\delta\sof{p_{x}\pm\delta_{x,z}+\sum\limits_{\nu}\ssof{k_{x,\nu}-k_{x-\hat{\nu},\nu}}}\,w\ssof{A_{x}+\delta_{x,z},B_{x},C_{x},p_{x};\kappa,s,s_{3},s_{4}}}\ ,\label{eq:partfderivpm}
\vspace{-5pt}\]
i.e. the derivative adds $+1$ to the first argument of the local weight $w$, and $\pm 1$ to the local delta function constraint for the site $z$ in \eqref{eq:partf4}, which can be interpreted as the effect of an insertion of an external positive or negative charge at that site. Analogously
\vspace{-5pt}\[
\frac{1}{\kappa}\partd{Z}{s_{z}^{3}}=\hspace{-6pt}\sum\limits_{\cof{k,l,\xi,\chi,p}}\bcof{\prod\limits_{x}\,K_{x}\of{\kappa}\,\e^{\ii\,\phi_{s}\,p_{x}}\,\e^{2\mu\,k_{x,4}}\,
\delta\sof{p_{x}+\sum\limits_{\nu}\ssof{k_{x,\nu}-k_{x-\hat{\nu},\nu}}}\,w\ssof{A_{x},B_{x}+\delta_{x,z},C_{x},p_{x};\kappa,s,s_{3},s_{4}}}\ ,\label{eq:partfderiv3}
\vspace{-5pt}\]
which adds $+1$ to the second argument of the local weight for site $z$, corresponding to the insertion of an external $\pi^{3}$ at that site.\\
In order to satisfy the delta-function or evenness constraint at site $z$ in \eqref{eq:partfderivpm} or \eqref{eq:partfderiv3} respectively, one either has to insert a second external field, or, if the corresponding sources $s$, $s_{3}$ are non-zero, to add an appropriate monomer. The former case would correspond to the start of a \emph{closed worm}\cite{Gattringer2} which samples a two-point function by moving one of the external fields around and updating flux-variables, while the latter case can be taken as the start of an \emph{open worm}\cite{Gattringer2}, which samples the condensate while moving the external field around.\\
By extending the closed worm by \emph{head-changing moves}, also mixed two-point functions like $\avof{\pi^{+}\of{x}\pi^{3}\of{y}}$ can be sampled during the worm, as explained in\cite{Rindlisbacher1}. Compared with the direct, computationally more expensive (in particular for large systems) measurement of correlators on configurations contributing to Z only, our method requires almost no additional computer time for these measurements. Furthermore, our method allows for a significant increase in statistics if the system is in the symmetry broken phase and the closing-time for the worm is of the order of the system size.
\vspace{-5pt}
\section{Results}\label{sec:results}\vspace{-5pt}
\subsection{Chiral symmetry breaking}
\begin{figure}[h]
\centering
\begin{minipage}[t]{0.55\linewidth}
\centering
\includegraphics[width=0.9\linewidth]{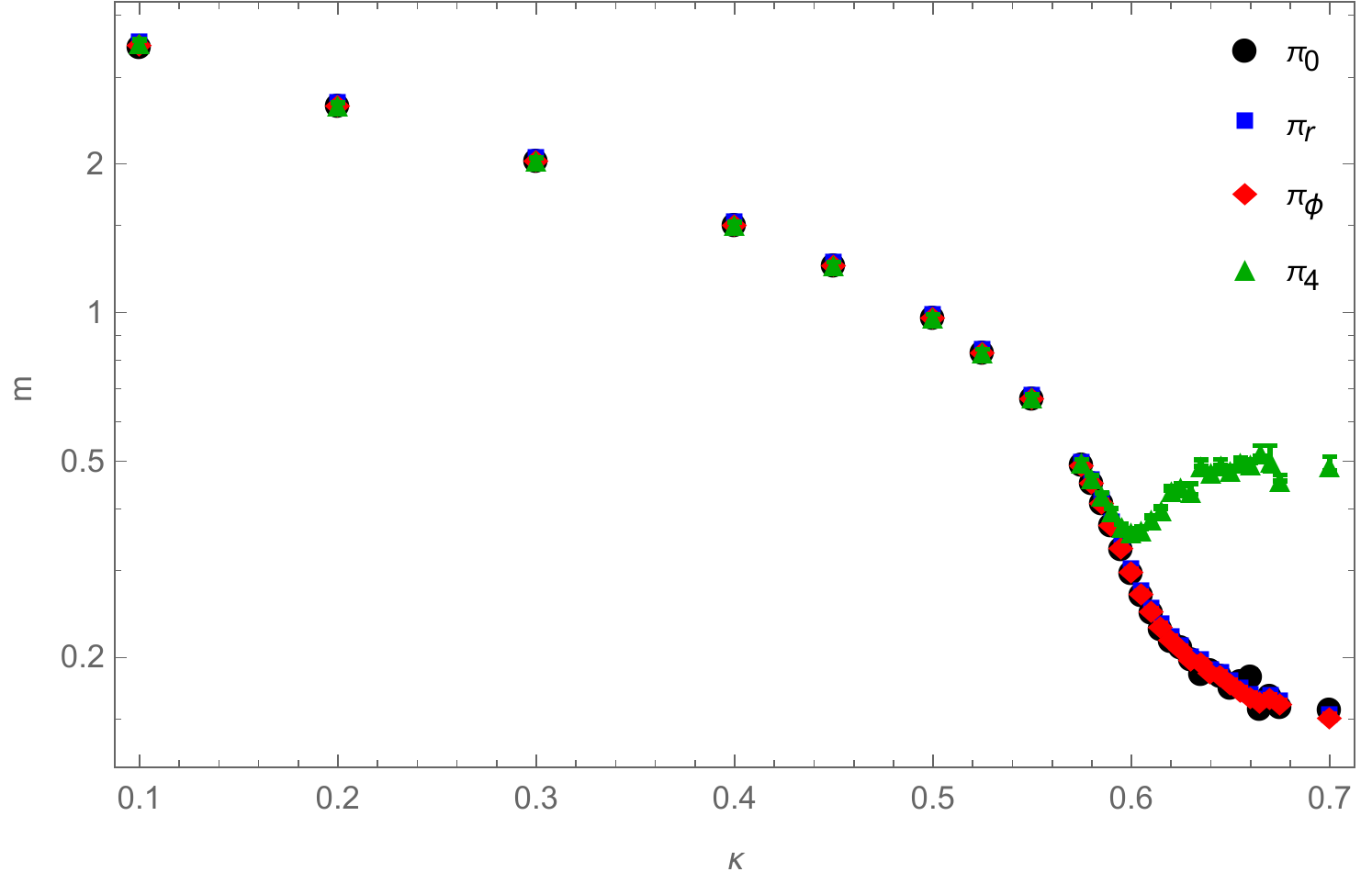}
\caption{\small Pion masses vs. $\kappa$ for a 4d system of size $8^3\times 12$ and a scalar source $s^{4}=0.01$. For $\kappa>\kappa_{pcr}\approx 0.605$ the $\pi^4$ mass increases again while the masses of the "true" pions decrease further as expected for pseudo-Goldstone bosons.}
\label{fig:chiralsb}
\end{minipage}\hfill
\begin{minipage}[t]{0.43\linewidth}
\centering
  \begin{tikzpicture}[scale=1]
  \path (0,0);
  \begin{scope}[shift={(0,1.6)}]
    \node (a) at (0,0) {};
    \node (b) at (0.75,0) {};
    \node (c) at (0,0.75) {};
    \node (d) at (0.704769,-0.256515) {};
    \node (e) at (0.256515,0.704769) {};
    \draw[color=white,fill=white,opacity=1] (a) circle (1.05);
    \draw[fill=black,opacity=0.2] (a) circle (1.05);
    \draw[color=white,fill=white,opacity=1] (a) circle (0.95);

    \draw[->,thick,color=red] ($(a)+(0,1)$) -- ($(b)+(0,1)$) node[below,pos=0.5]{$\pi_1$};
    \draw[->,thick,color=red] ($(a)+(0,1)$) -- ($(c)+(0,1)$) node[left,pos=0.5]{$\pi_2$};
    \draw[->,thick,color=red] ($(a)+(-1,0)$) -- ($(b)+(-1,0)$) node[below,pos=0.5]{$\pi_1$};
    \draw[->,thick,color=red] ($(a)+(-1,0)$) -- ($(c)+(-1,0)$) node[left,pos=0.5]{$\pi_2$};
    \draw[->,thick,color=red] ($(a)+(0.939693,-0.34202)$) -- ($(d)+(0.939693,-0.34202)$) node[below,pos=0.5]{$\pi_{r}$};
    \draw[->,thick,color=red] ($(a)+(0.939693,-0.34202)$) -- ($(e)+(0.939693,-0.34202)$) node[left,pos=0.5]{$\pi_{\phi}$};
  \end{scope}
  \end{tikzpicture}
\caption{\small Comparison of the $\of{\pi_1,\pi_2}$ basis, which mixes states under a $\Un{1}$ rotation and the $\of{\pi_{r},\pi_{\phi}}$ basis, which doesn't.}
\label{fig:diffbasis}
\end{minipage}
\end{figure}

At zero chemical potential and for vanishing sources $s$, $s^3$, $s^4$, the action \eqref{eq:latticeaction} has exact global chiral symmetry, which for sufficiently large $\kappa$, gets spontaneously broken to $\SU{2}$, where the pions play the role of the massless Nambu-Goldstone bosons. However, since in \eqref{eq:partf4} the global symmetry has been integrated out, together with the pion fields, we will need at least one non-vanishing source in the action in order to single out one of the degenerate vacua in the spontaneously broken phase. This is not only necessary to get non-zero expectation values for condensates, but also to obtain meaningful results for two-point functions. To understand this, note that what one calls a $\pi^1$, $\pi^2$, $\pi^3$ or $\pi^4$ excitation, is usually defined with respect to the direction (in the internal space) in which the symmetry is spontaneously broken. If this direction is not fixed, one cannot fix a coordinate system with respect to which one can define the different channels of the two-point function. Each choice of a fixed coordinate system would in this case lead to results for correlators in which the different channels are completely mixed or averaged together\footnote{The partition function sums over all degenerate vacua which all have different orientations in the fixed coordinate system with respect to which the two-point function is defined.}.\\
Specifying a non-zero source, e.g. a mass term $s_4$, breaks the chiral symmetry explicitly, but as long as the source is small enough that the change in the action caused by the source is not sufficient to compensate for the change in entropy that arises when the symmetry group is reduced to a subgroup, the effect of the source will essentially be invisible in the symmetric phase (this is also the case for chiral symmetry in QCD), as can be seen in Fig.~\ref{fig:chiralsb}, where the masses of the different excitations are shown as a function of $\kappa$: up to the pseudo-critical value $\kappa_{c}\approx 0.605$, the masses of all four channels are degenerate within errorbars. Only for $\kappa>\kappa_{c}$, when the action starts to dominate over entropy and the system develops long-range order, also the effect of the non-zero source term becomes visible: the mass of the $\pi^{4}$ (which, as we have chosen the non-vanishing source to be $s_{4}$, will be the excitation perpendicular to the manifold of degenerate vacua) increases again for increasing $\kappa$, while the other three pions continue to become lighter. However, due to the finite value of the scalar source $s_4$, the pions are now just pseudo-Goldstone bosons with finite masses. 

\subsection{Mass spectrum versus $\mu$}
\begin{figure}[h]
\centering
\begin{minipage}[t]{0.32\linewidth}
\centering
\includegraphics[width=\linewidth]{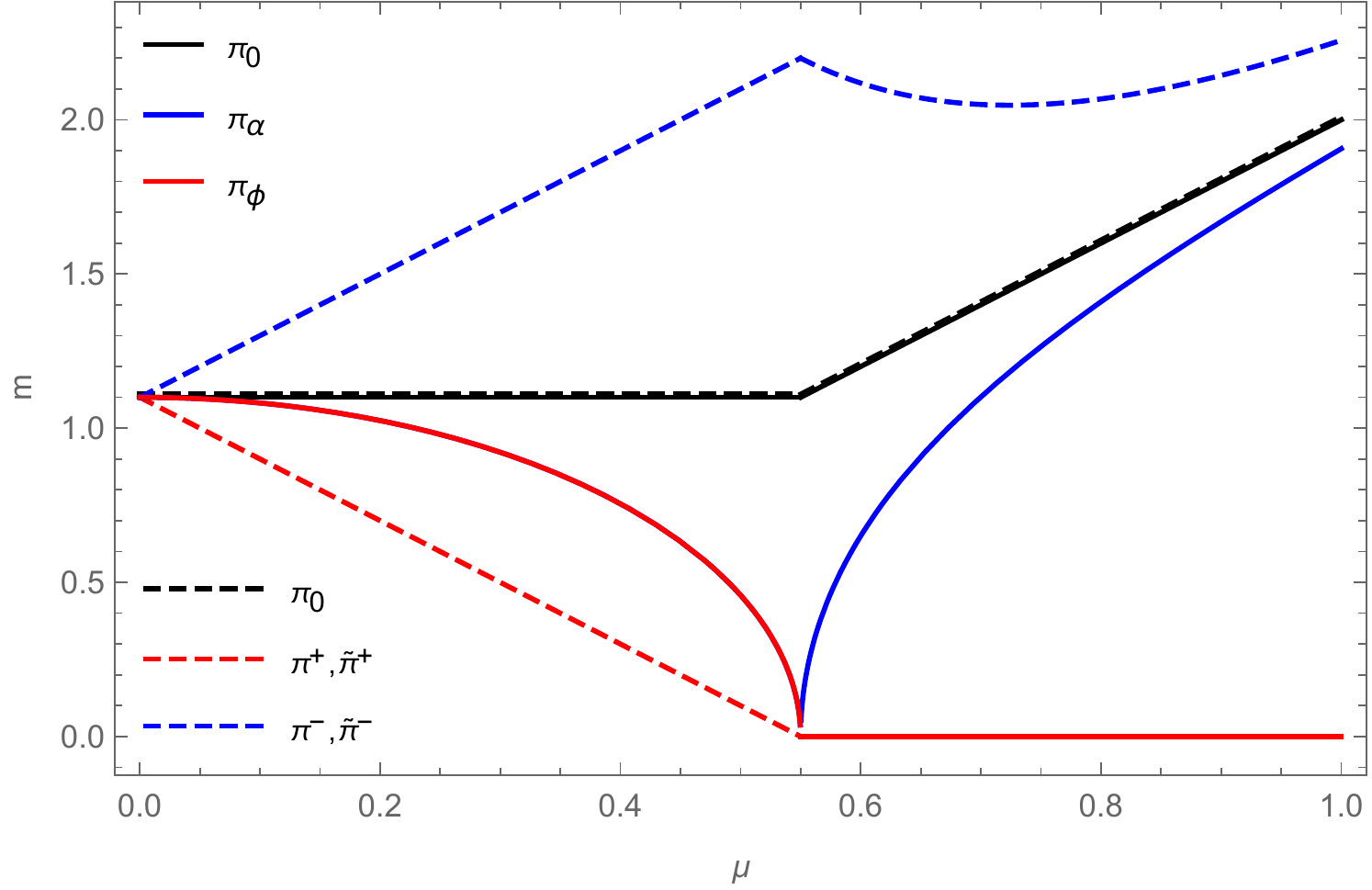}
\end{minipage}\hfill
\begin{minipage}[t]{0.32\linewidth}
\centering
\includegraphics[width=\linewidth]{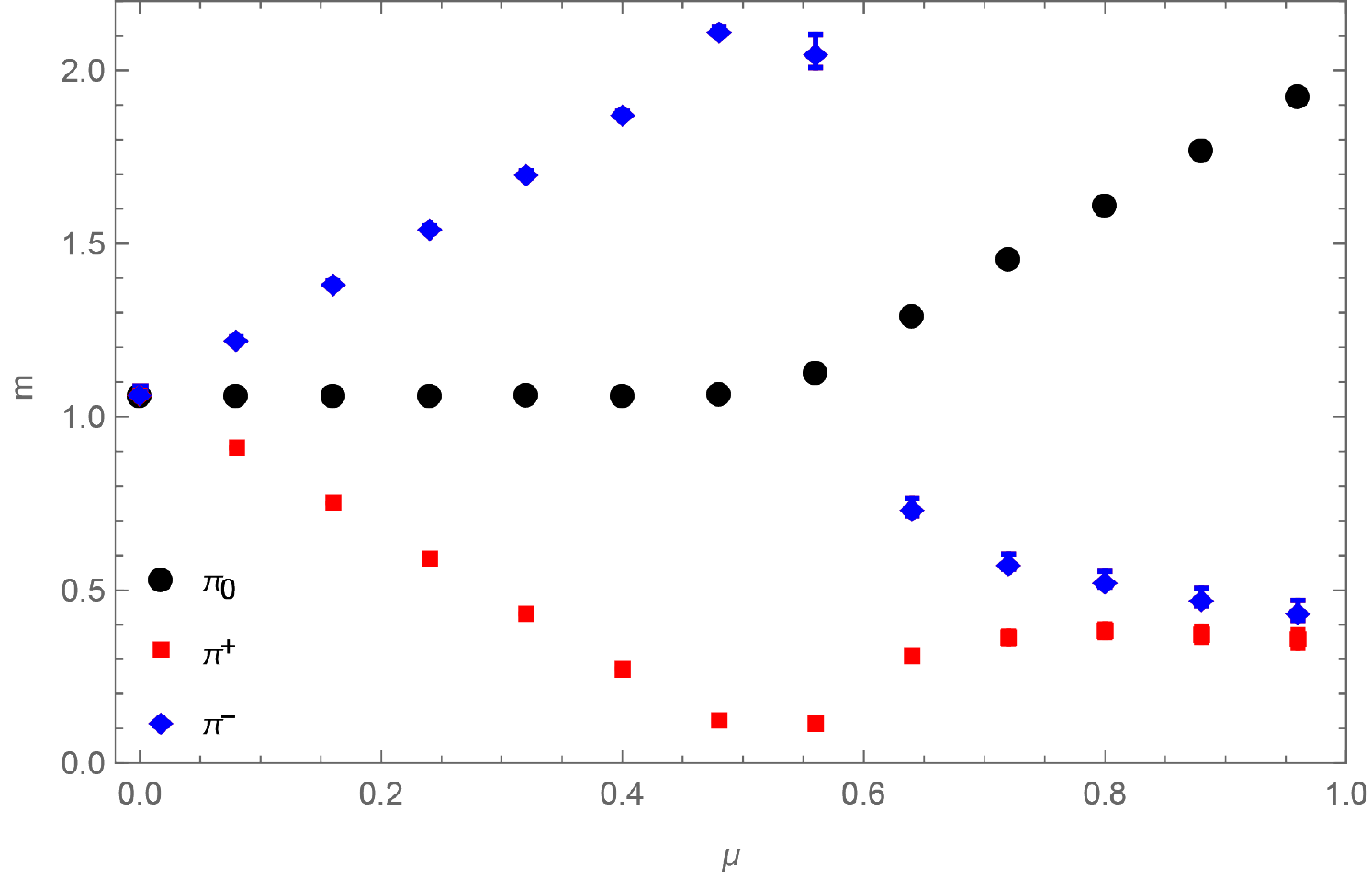}
\end{minipage}\hfill
\begin{minipage}[t]{0.32\linewidth}
\centering
\includegraphics[width=\linewidth]{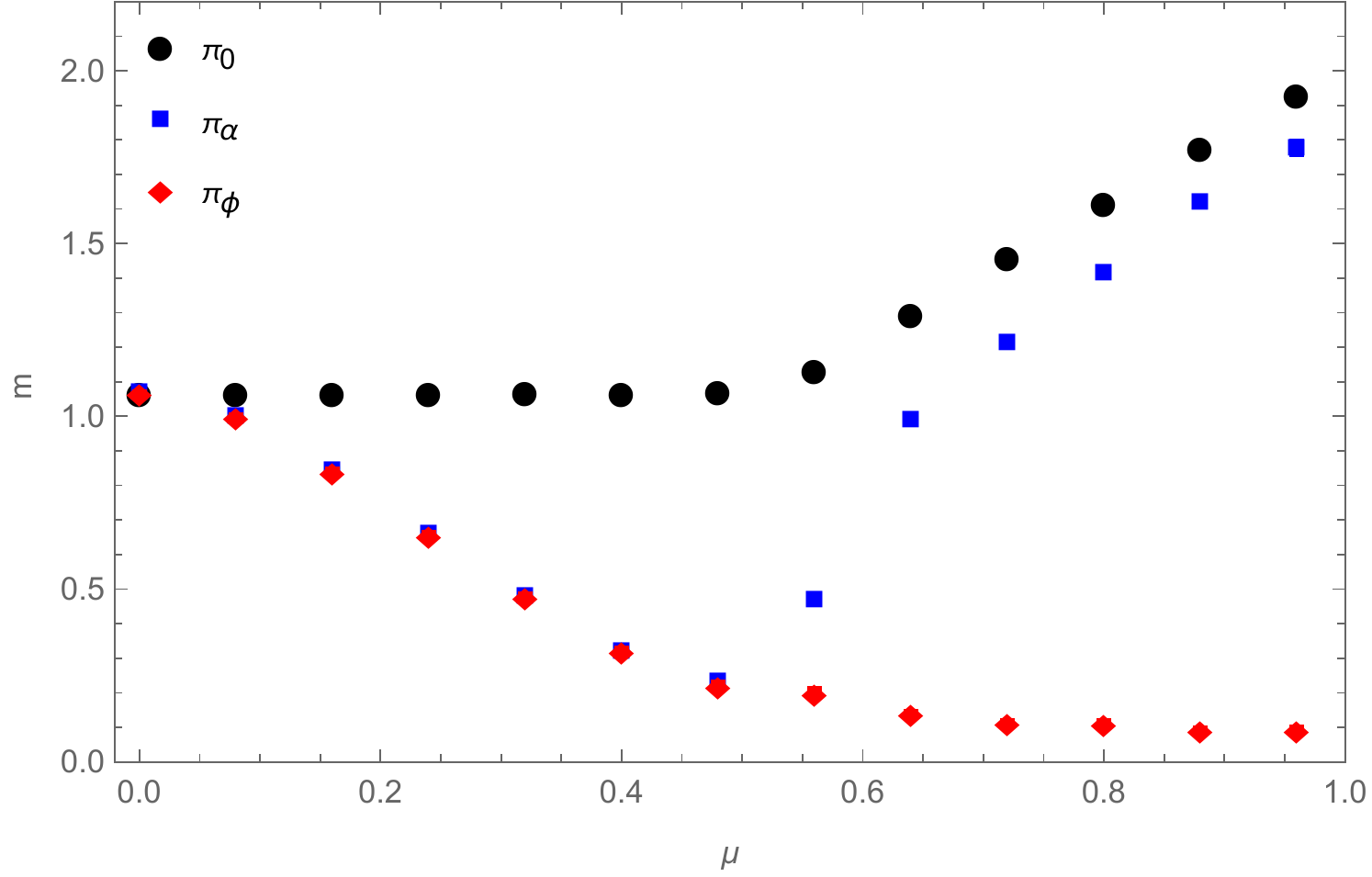}
\end{minipage}
\caption{\small Left figure: the dashed lines correspond to the masses of the three lightest excitations of the $\SU{2}$ principal chiral model, obtained from a saddle point calculation in the continuum theory (see. \cite{Son}). For $\mu<\mu_{cr}$ these are the masses of the three isospin-eigenstates $\pi_{0}=\pi_{3}$, $\pi^{+}$ and $\pi^{-}$, which for $\mu<\mu_{pcr}$ can also be directly measured in our lattice simulations (see middle figure). For $\mu>\mu_{cr}=m_{\pi}\of{\mu=0}/2$, isospin-charge is no longer a good quantum number: the ground state is no longer $\pi^{+}$ but $\tilde{\pi}^{+}$, a superposition of $\pi^{+}$ and $\pi^{-}$ and its mass can therefore not be measured straightforwardly in a corresponding lattice simulation. The same is true for $\tilde{\pi}^{-}$ which is also a superposition of the $\pi^{\pm}$. For $\mu>\mu_{cr}$ the masses of the \emph{angular excitations} $\pi_{\alpha}$ and $\pi_{\phi}$ (solid line in the left figure, obtained from the Hessian of the effective potential at the saddle point) are better suited to be determined on the lattice, as shown in the right figure. However, as explained in the text, in order to obtain the expected behavior for the blue data, one has to adjust the angle $\alpha_{s}$ as a function of $\mu$ when expressing \eqref{eq:angularcorr} in terms of \eqref{eq:stdcorr}.}
\label{fig:isomassspec}
\end{figure}

For finite scalar source $s_4$ and sufficiently large $\kappa$, the model can be interpreted as a low-energy effective model for two-flavor QCD with finite, degenerate quark masses, coupled to an isospin chemical potential $\mu$. For $\mu>0$, the $\SU{2}$ symmetry is explicitly broken to an isospin-$\Un{1}$ symmetry, and if the value of the chemical potential exceeds half the value of the pion mass, the chiral condensate (which we have chosen to be in the $\pi^4$-direction) starts to rotate into a pion-condensate and thereby spontaneously breaks also this remnant $\Un{1}$ symmetry.\\
As pointed out at the end of Sec. \ref{ssec:partf}, the partition function \eqref{eq:partf4} is completely independent of $\phi_{s}$, and it is therefore not possible to break the isospin-$\Un{1}$ symmetry explicitly by fixing $\phi_{s}$. However, we don't really need to break the symmetry, we only need to be able to define a coordinate system in which the physically distinct excitations have well defined, fixed orientations in all degenerate vacua (see Fig.~\ref{fig:diffbasis}). This can be achieved by using polar coordinates $\pi_{r}$ and $\pi_{\phi}$ in the $\pi^1$-$\pi^2$ plane, and specifying a non-zero value for $s$ to shift the projection of the dominant vacuum into the $\pi^1$-$\pi^2$ plane slightly off the origin in that plane where the polar coordinates would be singular.\\
As $\SU{2}\sim \Sph{3}\in \mathbb{R}^{4}$, we can think of $\of{\bar{\pi},\pi_{4}}$ as coordinates in $\mathbb{R}^{4}$ (spanned by the basis $\of{\bar{\sigma},\id}$). If the chiral symmetry is broken by a non-zero source $s_{4}$, leading to a vacuum expectation value for $\pi_{4}$, the dominant vacuum sits at the "north pole" of $\SU{2}$, the identity $\id$. There the tangent space is spanned by $\bar{\sigma}$ and its orthogonal complement by $\id$. The fluctuations in $\bar{\pi}$, the pions, are therefore tangential to $\SU{2}$ as it should be. But as soon as the vacuum rotates away from the identity, this is no longer the case and tangential excitations will in general be superpositions of fluctuations in $\bar{\pi}$ and $\pi_{4}$, which are most easily expressed in terms of spherical coordinates $\pi_{\alpha}$, $\pi_{\theta}$ and $\pi_{\phi}$. By defining in addition to $\phi_{s}$ also the spherical sources $\alpha_{s}$ and $\theta_{s}$, such that $s_{4}=\tilde{s}\cos\of{\alpha_{s}}\ ,\ s_{3}=\tilde{s}\sin\of{\alpha_{s}}\cos\of{\theta_{s}}$ and $s=\tilde{s}\sin\of{\alpha_{s}}\sin\of{\theta_{s}}$, we can define
\vspace{-2pt}\[
\frac{1}{\of{\kappa\,\tilde{s}}^2}\spartd{\log\of{Z}}{\alpha_{s,x}}{\alpha_{s,y}}\quad,\quad \frac{1}{\of{\kappa\,s}^2}\spartd{\log\of{Z}}{\phi_{s,x}}{\phi_{s,y}}\quad \text{and}\quad \frac{1}{\of{\kappa\,\tilde{s}\,\sin\of{\alpha_{s}}}^2}\spartd{\log\of{Z}}{\theta_{s,x}}{\theta_{s,y}}\ ,\label{eq:angularcorr}
\vspace{-2pt}\]
which can be expressed via the chain rule in terms of the directly measurable two-point functions,
\vspace{-2pt}\[
\frac{1}{\kappa^2}\spartd{\log\of{Z}}{s_{x}}{s_{y}}\quad,\quad\frac{1}{\of{\kappa\,s}^{2}}\spartd{\log\of{Z}}{\phi_{s,x}}{\phi_{s,y}}\quad,\quad\frac{1}{\kappa^2}\spartd{\log\of{Z}}{s_{3,x}}{s_{3,y}}\quad\text{and}\quad\frac{1}{\kappa^2}\spartd{\log\of{Z}}{s_{4,x}}{s_{4,y}}\ .\label{eq:stdcorr}
\vspace{-2pt}\]
However, in order to reproduce with the data from our simulations the masses of $\pi_{\alpha}$, $\pi_{\theta}=\pi_{0}$ and $\pi_{\phi}$ vs. $\mu$ as obtained from the Hessian of the effective potential of the continuum theory, evaluated at the saddle point (see solid lines in Fig. \ref{fig:isomassspec}, left), one has to determine $\alpha_{s}$ in the expressions for \eqref{eq:angularcorr} (in terms of \eqref{eq:stdcorr}) not from the values of $s$ and $s_4$ (we set $\theta_{s}=\pi/2$) used in the simulation, but instead from the values of the measured condensates, i.e. $\alpha_{s}=\arctan\ssof{\avof{\pi_{r}}/\avof{\pi^{4}}}$ (see Fig. \ref{fig:isomassspec}, right). This suggests that also during the simulation, the sources $s$ and $s_{4}$ should be adjusted such that $s/s_{4}=\avof{\pi_{r}}/\avof{\pi^{4}}$, which changes as a function of $\mu$.\\

\vspace{-5pt}
\section{Summary and outlook}\label{sec:conclude}\vspace{-5pt}
We have derived a sign-problem free lattice formulation for the $\SU{2}$ principal chiral model coupled to a chemical potential and including source terms. Furthermore, we have sketched a method to efficiently measure general one- and two-point functions during the worm updates, to be explained in more detail in \cite{Rindlisbacher1}.\\
The above was then used to investigate the difficulties associated with the study of spontaneous symmetry breaking in lattice formulations where the symmetry that should break is integrated out exactly as in lattice QCD.\\

\end{document}